# Phase Transitions of Chiral Spin Textures via Dipolar Coupling in Multilayered Films with Interfacial Dzyaloshinskii-Moriya interactions

Javier F. Pulecio[1,2], Aleš Hrabec[3,4], Katharina Zeissler[3], Yimei Zhu[1], and Christopher H. Marrows[3]

[1]Department of Condensed Matter Physics, Brookhaven National Laboratory, Upton, New York 11973, USA

[2]National Institute of Standards and Technology, Boulder, Colorado 80305, USA

[3]School of Physics and Astronomy, University of Leeds, Leeds LS2 9JT, United Kingdom.

[4]Present Address: Laboratoire de Physique des Solides, CNRS, Universités Paris-Sud et Paris-Saclay, 91405 Orsay Cedex, France

Under the correct conditions, Dzyaloshinskii-Moriya interactions (DMI) can lead to topologically protected spin textures such as skyrmions. The application of DMI for spin-based computation and memory technologies is promising and requires a detailed consideration of the role intrinsic energies have on stabilizing the diverse textures observed thus far. Here we experimentally investigate the effect of dipolar energy from interlayer coupling on the remanent spin textures found at room temperature for interfacial DMI multilayers of [ Pt \ Co \ Ir ]$_{\times N}$. The total dipolar energy is modified by increasing the number of layer repetitions $N$ which result in different phases of chiral magnetic textures. We use the phase transitions to estimate the DMI energy present in the ultrathin films to be $D$ = 2.1 ± 0.1 mJ/m$^2$. Unlike traditional perpendicular magnetic anisotropy films, these multilayer films with finite DMI exhibit isolated hedgehog skyrmion bubbles as well as sub 100 nm labyrinth domains with cycloidal homochiral Néel walls which are dependent on DMI as well as the total dipolar energy.

Recent observations of the magnetic topology [1] found in thin magnetic films that exhibit the interfacial Dzyaloshinskii-Moriya interaction (DMI) have shown a variety of topologically non-trivial spin textures [2], [3], with out-of-plane (OOP) domain sizes ranging from several microns down to sub 100 nm skyrmion bubbles [4]–[9]. While the diverse nature of the observed textures is related to the perpendicular magnetic anisotropy (PMA) and DMI strength, the effect of interlayer dipolar coupling on the remanent states has yet to be explored in great detail. A cause for the different spin configurations observed thus far in interfacial DMI systems, stems from the dissimilar materials systems. This is further confounded by the various number of ferromagnetic (FM) stack repetitions used, as it is well known that interlayer dipolar coupling found in PMA materials can affect the remanent magnetic states [10], [11]. Traditional PMA films without DMI prefer Bloch-type domain walls in order to minimize magnetostatic energy, and so, nucleate spin textures with trivial topology. The non-trivial chiral spin textures found in DMI materials are currently of great interest due to the propensity to exhibit unique topological phenomena such as protection/stabilization, pinning mitigation, and Hall effects [1], [2], [12]–[14]. Understanding how to nucleate and control chiral textures, via current driven dynamics for example, could help unlock the potential of novel spintronic technologies [15]–[17].

In ultrathin films with interfacial DMI, Néel type walls are stabilized over Bloch walls since the DMI acts as an effective field normal to the plane of the wall [18]. Moreover, the handedness of the DMI enforces a single chirality (i.e. homochiral) for all of these Néel walls, differentiating them from the aforementioned traditional PMA films where the DMI is not substantial. In the following we use Lorentz TEM and full 3-D micromagnetic simulations to investigate the magnetic phases that exist post demagnetization as a function of the number of stack repetitions $N$. The Lorentz transmission microscopy (LTEM) experiments allowed for the determination of the domain wall type and the magnetization direction of the domain, whilst the micromagnetic simulations were used to model the effect of the interlayer dipolar coupling and DMI strength $D$ on the magnetic phases. Fully 3-D micromagnetic models were employed to study the chiral spin textures, which we have found can lead to different ground states than those implemented with effective medium generalizations since coupling between discrete magnetic layers is treated more rigorously. We demonstrate how tuning the dipolar energy via the number of stack repetitions $N$ for ultrathin films with interfacial DMI causes the system to relax into distinct phases of magnetic chiral textures and discuss how these transitions can be used to estimate $D$.

**Ultrathin multilayered films with interfacial DMI –** Figure 1 (a) shows a schematic for two stack repetitions ($N$ = 2) of Pt 1.4 nm \ Co 0.8 nm\ Ir 0.4 nm. The interlayer dipolar coupling occurs between the ferromagnetic cobalt layers, where the total dipolar coupling energy increases as a function of higher number of repetitions as the total volume of magnetic material increases. The ultrathin magnetic film systems of [ Pt 1.4 nm \ Co 0.8 nm \ Ir 0.4 nm ]$_{\times N}$ were deposited by dc magnetron sputtering onto Si$_3$N$_4$ membrane substrates in a vacuum chamber with a base pressure of $10^{-8}$ Torr. Figure 1 (b-c) shows a $M(H)$ hysteresis loop and the temperature dependence of the saturation magnetization $M_{sat}(T)$, obtained with a superconducting quantum inference device vibrating sample magnetometer (SQUID-VSM), which was used to determine the magnetic properties of the films. Representative measured magnetic parameters for these stacks are $M_{sat}$ = 0.9 $\pm$ 0.2 MA m$^{-1}$, where only the volume of the FM Co layers was used to calculate the magnetization, an effective perpendicular anisotropy value of $K_{eff}$ = 0.35 $\pm$ 0.03 MJ m$^{-3}$, and exchange stiffness coefficient of $A$ = 11 $\pm$ 1 pJ m$^{-1}$ determined by temperature dependent magnetization measurements in Figure 1 (c).

**Lorentz TEM of spin textures –** Figure 2 (a-b) shows illustrative Néel wall profiles and the simulated Fresnel mode LTEM images that arise from them. At normal incidence of the electron beam (zero tilt angle θ) there is no magnetic contrast that arises from LTEM for such a wall, since the only in-plane (IP) magnetization that can cause a phase shift of the electron wave occurs at the wall [19], and results in a deflection along the length of the wall. When the sample is tilted (in this case θ = 15°), a portion of the OOP domain magnetization is tilted IP and leads to contrast that can provide the position of the Néel domain wall [2]. Once the film is saturated into a single domain state, as in Figure 2(c), the sample's magnetization is uniform, which naturally results in there being no LTEM contrast.

Figures 2 (d-f) are room temperature experimental Lorentz TEM images for $N$ = 5. The dark particle in the lower right corner is debris and was used to provide a position reference. At zero tilt (Figure 2(d)) there is no magnetic contrast present, which signifies the lack of any Bloch type walls that would result in magnetic contrast at zero tilt from the wall itself. When the sample is then tilted by 15° (Figure 2(e)), black and white walls appear, indicative of the presence of Néel type walls. To ensure the wall contrast was magnetic in nature, a saturating field of $B_z$ = 50 mT was applied which annihilated the Néel walls as shown in Figure 2 (f). Prior to annihilation, an external field of $B_z$ = 48 mT was applied such that the OOP domains were no longer distinguishable and the domain walls were adjacent. This configuration was stabilized by the topological charges of the homochiral Néel walls which provided an energy barrier before annihilation [2].

**The effect of dipolar coupling** – Figure 3 shows LTEM images of the remanent magnetic states post demagnetization for different number of stack repetitions. The films where initially demagnetized *ex-situ* using a OOP AC demagnetization process with a max starting field of 1 T, which repeatedly brought the system into the representative global states depicted in Figure 3. The remanent state for a two stack repetition ($N$ = 2) multilayer, imaged at zero field in the LTEM, is shown in Figure 3 (a). The ultrathin films of $N$ = 2 show a propensity for large domains that extend hundreds of microns beyond the field of view in (a). The image also shows a few isolated skyrmion bubbles indicated by the red arrows. These bubble domains are enclosed by the homochiral Néel walls discussed in the previous section, making them topologically non-trivial [1] and of the hedgehog skyrmion variety [20].

Figure 3(b) shows a $N$ = 5 system, where homochiral Néel walls enclose large domains that are on average smaller than the domains found in the $N$ = 2 system. Figure 3(c) is a $N$ = 10 system with very dense OOP labyrinth domain patterns where the wall density has increased dramatically and sub 100 nm domain widths are evident. We attribute the variation of the remanent spin textures to the stronger dipolar coupling that arises from a larger total volume of magnetic material as the number of repetitions is increased.

The buckled stripe (i.e. labyrinth) domains in Figure 3 (c) are delimited by homochiral Néel walls and have a cycloidal topology [18], [21], [22]. We will refer to this spin configuration as the cycloidal state henceforth. We note an 'ideal' magnetic cycloid has the same topology but would not be buckled (ie straight stripes) and have a constant angular velocity in 1D, such that the OOP domain width equals the domain wall width. This is similar to the nomenclature used in literature for magnetic skyrmions with extend bubble domains, where the topology is used to define the skyrmion even though the bubble domain width is much larger than the wall width.

**Simulations of Interlayer Dipolar Coupling** – Figure 4 shows micromagnetic simulations of a disc where the fundamental stack consisted of Pt 1.6 nm \ Co 0.8 nm \ Ir 0.4 nm with a cell size of 2 x 2 x 0.8 nm$^3$ and a disc diameter of $d_d$ = 500 nm. This was carried out in the OOMMF [23] with an additional DMI energy term of the interfacial symmetry form [22]. The top and bottom rows show $N$ = 1 and $N$ = 2 stack repetitions, respectively, for different values of the DMI constant $D$. The simulations shown in Figure 4 were initialized with a bubble domain diameter of $d_b$ = 300 nm and allowed to relax in zero field. The simulations were also initialized with an initial skyrmion bubble size of $d_b$ = 30 nm to ensure that the initial conditions did not affect the remanent states, which they did not.

Focusing on $N$ = 1 (top row) in Figure 4, the phase transitions of the magnetic states as a function of the DMI energy are evident. In the absence of DMI $D$ = 0 mJ m$^{-2}$, the hedgehog skyrmion bubble collapses onto itself leaving a single domain state. For $D$ = 2.3 mJ m$^{-2}$ the bubble appears to be stable at a diameter of $d_b$ = 248 nm. For $D$ = 2.4 mJ m$^{-2}$ the domain is no longer circular and is the precursor for the cycloidal state. For values of $D$ greater than or equal to 2.5 mJ m$^{-2}$ the system minimizes the total energy by reducing the stray field energy (i.e. domain size) and increasing the domain wall density into a cycloidal state where there is a dense array of homochiral domain walls, each of which is a half-turn of a cycloid. This can take place since stronger $D$ reduces the energy cost of a domain wall to zero [12]. The homochiral Néel walls and OOP domains form a cycloidal state where the length of the full cycloidal revolution, $l_c$, reduces as the DMI energy $D$ increases. In Figure 4, we highlight the phase transition into the cycloidal state due to the DMI energy as $t_1$ (magenta) and will refer to this critical DMI energy as $D_c$.

For $N$ = 2, similar transitions and phases exist as those found in $N$ = 1, but the total dipolar energy of the system is greater due to the additional FM, and the two layers are interlayer dipolar coupled. As can be seen in the bottom row of Figure 4, the dipolar coupling energy reduces the DMI energy necessary for the element to make the transition into a cycloidal state from $D$ = 2.5 mJ m$^{-2}$ for $N$ = 1, to $D$ = 2.4 mJ m$^{-2}$ for $N$ = 2 as detailed by the $t_1$ magenta label. Interestingly, another type of transition into the cycloidal state occurs due to the increased dipolar energy indicated by $t_2$ (black). This suggests that modifying the number of layer repetitions and observing a $t_2$ transition into the cycloidal state, due to the interlayer dipolar coupling, may provide a useful estimate of the DMI energy $D$ in an experiment, and will be discussed in greater detail in the subsequent section. A subtler effect from interlayer dipolar coupling can be seen for $l_c$ in Figure 4, where an increase in dipolar energy effectively reduces the length of the full cycloidal revolution, for example at $D$ = 3.0 mJ m$^{-2}$, by shrinking the labyrinth domain widths. Conversely, in the skyrmion bubble state, the dipolar energy has the opposite effect, increasing the bubble domain size as the dipolar energy is increased.

Experimentally modifying and quantifying the DMI energy in these systems, similar to the simulations in Figure 4, is currently a critical area of research, and so, inducing a phase transition via $t_1$ could allow for the approximation of $D$. An alternate approach is to modify the dipolar energy by changing $N$ to better understand $D$ via $t_2$. For the simulated value of $D$ = 2.4 mJ m$^{-2}$ and $N$ = 1 we see the system can transition via $t_2$ (i.e. increased dipolar energy) into a cycloidal state by changing the number of repetitions to $N$ = 2. Similarly, in Figure 3 we experimentally observe the system transitions from large domains for $N \leq 5$ to the cycloidal state for $N$ = 10. In the following we discuss the experimentally observed phase transitions in Figure 3 as $t_2$ type transitions resulting from increased interlayer dipolar coupling energy and correlate it to the DMI energy.

**DMI and interlayer dipolar coupling energy –** As mentioned earlier, the DMI allows for Néel walls to stabilize in conventional PMA systems which would otherwise prefer Bloch type walls. The critical DMI energy necessary to stabilize Néel walls is $D_N = \frac{4}{\pi}\sqrt{A/K_{eff}}\ K_D$, where $A$ is the exchange constant, $K_{eff}$ is the effective OOP uniaxial anisotropy, and $K_D$ is the demagnetization (i.e. shape) anisotropy of the wall [18]. For our measured parameters, the critical value needed to stabilize a Néel wall in a single stack repetition is $D_N$ = 0.135 mJ m$^2$ and Figure 5(a) shows how $D_N$ is affected by $A$ and $K_{eff}$ for a single repetition ($N$ = 1). This sets an approximate lower bound on the DMI energy as Néel walls were identified for $N$ = 2, but as we will discuss next, can be up an order of magnitude lower than the actual DMI energy.

The simulations in Figure 4 demonstrated a transition from magnetic phases with high orders of symmetry, such as a bubble, into the cycloidal phase. For $N$ = 1, we can predict this transition via $t_1$ using an analytical expression of the critical DMI energy for the cycloidal state (when the DMI reduces the DW energy to zero) as $D_c = \frac{4}{\pi}\sqrt{A\ K_{eff}}$, where $A$ is the exchange constant and $K_{eff}$ is the effective OOP uniaxial anisotropy [21]. For our films, the analytically predicted value necessary to observe a cycloidal state, when $N$ = 1, is $D_c$ = 2.5 mJ m$^{-2}$ and detailed in Figure 5 (b). This is approximately 25 times larger than the critical DMI value necessary to stabilize Néel walls $D_N$. We can apply this value as an approximate upper bound for our systems. While we did observe Néel wall stabilization for $N$ = 2, we did not observe the cycloidal state. This would suggest the DMI energy in our system lies between $D$= 0.135 - 2.5 mJ m$^{-2}$, as is typical for multilayer systems of this sort [5], [24].

It is worth noting the models are quite sensitive to the values of $A$ as well as $K_{eff}$, so careful attention should be placed on these measurements. When measuring $M_s$, as shown in Figure 1 (b), proximity magnetization can be a factor in calculating $M_s$ and is directly related to the effective anisotropy by $K_{eff} = \frac{1}{2}M_s(H_{eff} + \mu_0 M_s)$, where $H_{eff}$ is the effective anisotropy field and $\mu_0$ is the permeability of free space. The determination of $A$ via the temperature dependent measurement used in Figure 1 (c) can also have complexities related to the magnon scattering process and can also introduce offsets. As detailed Figure 5 (b) for N =1, if we keep $K_{eff}$ = 0.35 ± 0.03 MJ m$^{-3}$ and increase the exchange by 50% to $A$ = 16.0 pJ m$^{-1}$, the critical DMI energy for the cycloidal state of would also increase by 50% to $D_c$ = 3.0 mJ m$^{-2}$.

Figure 5 (c) shows a micromagnetic (numeric) simulated phase diagram of $N$ and $D$ using the experimentally determined magnetic parameters $A$, $K_{eff}$, and $M_s$ of our films. By changing the number of layers $N$ and the DMI energy $D$, we can construct a plot detailing the effect of interlayer dipolar coupling and the associated stable phases of a hedgehog skyrmion state (magenta circles), a transitional state (green circles), and a cycloidal state (light blue circles). The general spin configurations of the stabilized states are shown in Figure 4, and are also labeled with colored circles in the upper left corners for reference. The blue box in Figure 5 (c) outlines in the single repetition case ($N$ =1) where the $t_1$ transition into the cycloidal state due to higher DMI energy occurs at $D_c$= 2.5 mJ m$^{-2}$. Note, the numeric and analytical models agree on the predicted DMI energy necessary to stabilize the cycloidal state $D_c$ and is highlighted by the star marker. The phase diagram then allows us to further understand $t_2$ phase transitions which increase the systems total dipolar energy as a function of layer repetitions.

In our experiments, we observed the transition into the cycloidal state occur at $N$ = 10. Following the simulated phase transitions in Figure 5 (c), which agrees well with the analytical model for $N$ = 1, this would seem to suggest an approximate minimum DMI energy of $D \approx$ 2.0 mJ m$^{-2}$ as outline by the red box overlay. The fact that we did not observe the cycloidal state at $N$ = 5, sets the upper bound of $D \approx$ 2.2 mJ m$^{-2}$. The simulated phases in this DMI energy regime also agree well with our observations of $N$ =2 (large domains and skyrmion bubbles) and $N$ = 5 (no skyrmion bubbles and smaller domain widths) and $N$ = 10 (the cycloidal state). Notably, the stability of isolated hedgehog skyrmion bubbles in these systems occurs before the system goes into the cycloidal state as outlined in Figure 5 (c). This agrees well with our experiments shown in Figure 3(c), where the combination of the total dipolar and DMI energies is sufficiently low to stabilize the hedgehog skyrmions for *N* =2.

## Conclusion

In conclusion we have shown interlayer dipolar coupling can be tuned to stabilize different remanent chiral spin textures found in multilayered film stacks of [ Pt \ Co \ Ir ]$_{×N}$ with interfacial DMI. Using Lorentz TEM imaging, we determine the presences of out-of-plane magnetized domains with homochiral Néel type walls that are stabilized by the DMI energy. The increased dipolar energy, that results from greater number of layer repetitions, transitioned the system from large single domain states for $N$ = 2 to cycloidal labyrinth domains for $N$ = 10. Using the experimentally measured magnetic parameters, and correlating the simulated phase transitions into the cycloidal state induced by interlayer dipolar coupling, we estimate a DMI energy of $D$ = 2.1 ± 0.1 mJ m$^{-2}$ in these ultrathin magnetic heterostructures with interfacial DMI. Importantly, our Lorentz TEM observations agree with simulations that demonstrate the stability of hedgehog skyrmion bubbles occur just prior to the cycloidal phase and is directly related to the interlayer dipolar coupling. As the number of stack repetitions are increased, so is the total dipolar energy of the system which reduces the critical DMI energy need for the cycloidal state ($D_c$). Therefore, the stabilization of hedgehog skyrmions in these systems is directly related to the DMI energy as well the total dipolar energy.

## Literature

[1]    N. Nagaosa and Y. Tokura, "Topological properties and dynamics of magnetic skyrmions," *Nat.*

of skyrmions in magnetic nanodisks without the Dzyaloshinsky-Moriya interaction," *Phys. Rev. B*, vol. 88, no. 5, p. 54403, Aug. 2013.

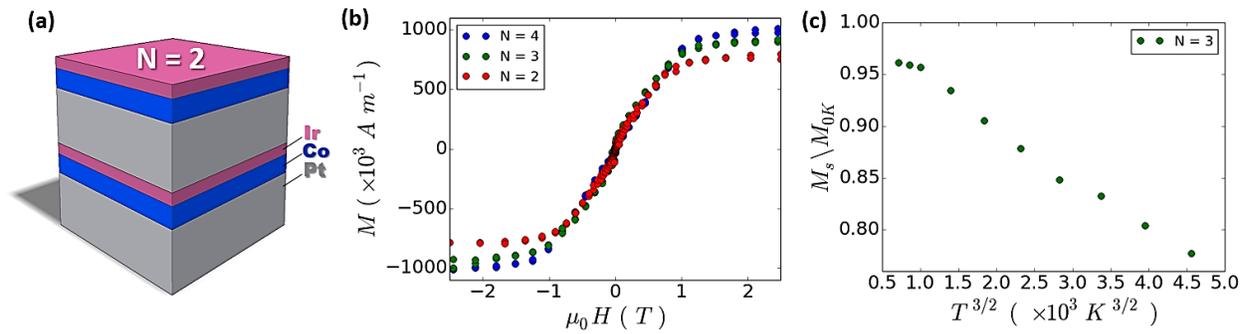

Figure 1 – Sample Geometry and Magnetometry. (a) Two repetitions (N=2) of an ultrathin multilayered stack of [ Pt 1.4 nm \ Co 0.8 nm \ Ir 0.4 nm ]. SQUID-VSM $M(H)$ curves for $N$ = 2, 3, 4 multilayers (b) and $M(T)$ plot for $N$ = 3 (c).

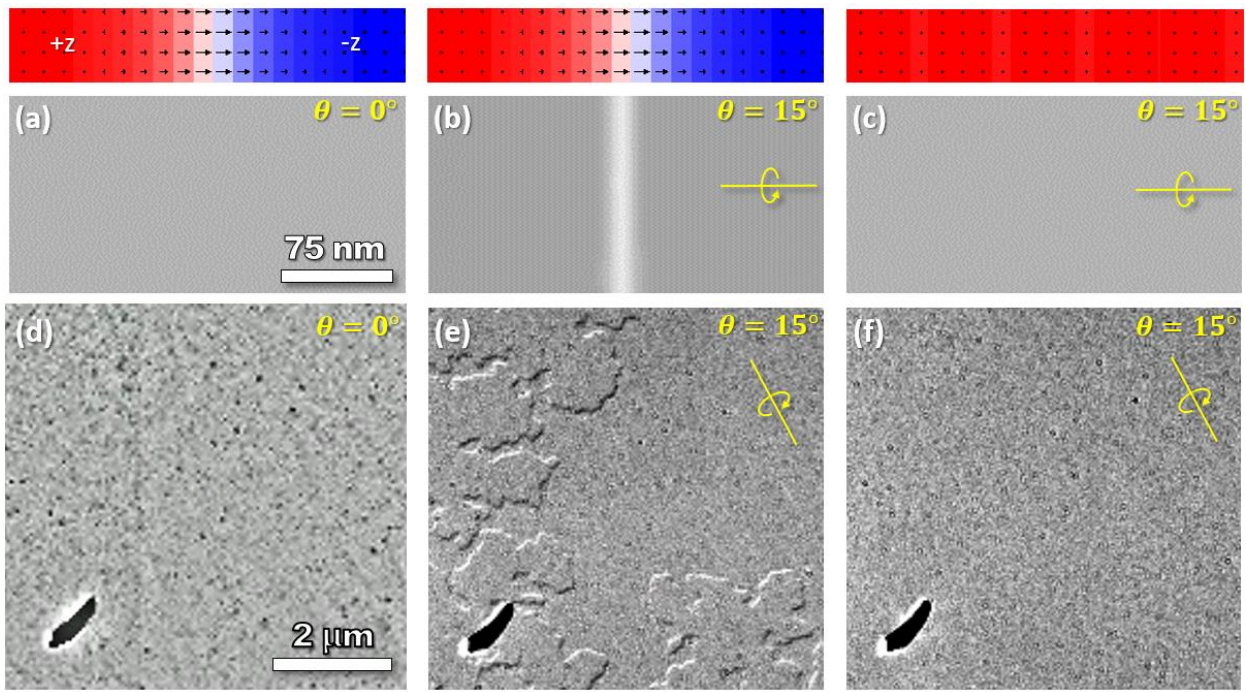

Figure 2 – LTEM Imaging of homochiral Néel walls. (a-b) Zero field LTEM image simulations for different tilts of a Néel wall between out of plane domains. At zero tilt $\theta = 0°$ (a) there is no magnetic contrast present, while tilted $\theta = 15°$ magnetic contrast reveals the position of the Néel walls. The film was saturated in (c) into a single domain state and, as expected, the LTEM image simulations exhibits no magnetic contrast. (d-e) Zero field experimental LTEM images at different tilts. No magnetic contrast is present in (d), but tilted 15° confirms the presence of Néel walls (e). The film was then saturated with a 50 mT out-of-plane field and the Néel walls were annihilated (f).

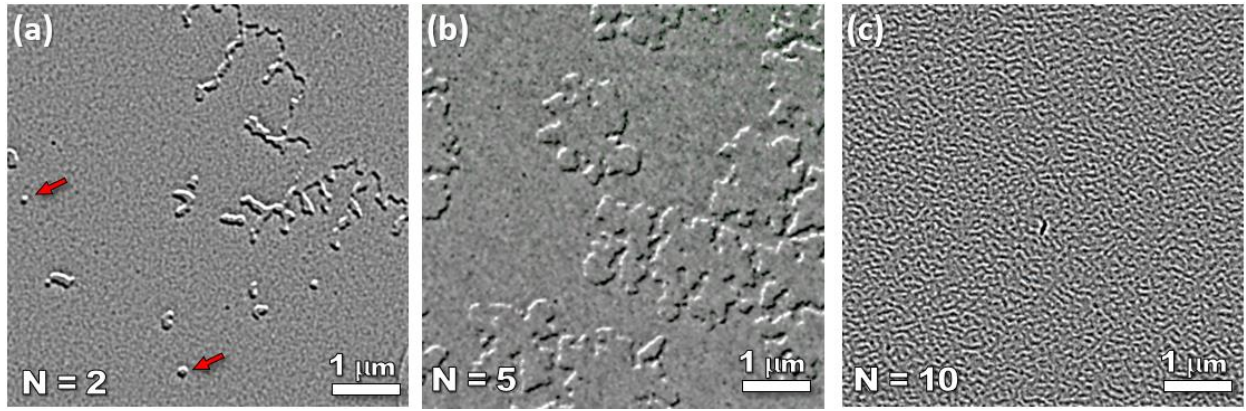

Figure 3 - Lorentz TEM images of demagnetized [ Pt \ Co \ Ir ] stacks of $N$ = 2 (a), $N$ = 5 (b), and $N$ = 10 (c) repetitions. The black and white line contrast represent Néel domain walls which appear at a tilt of $\theta = 15°$. The red arrows in (a) identify isolated skyrmion bubbles. From (a) to (b) the wall density increases while the average domain size decreases as the interlayer dipolar coupling becomes stronger for larger $N$. In (c) the system transitions into a cycloidal phase.

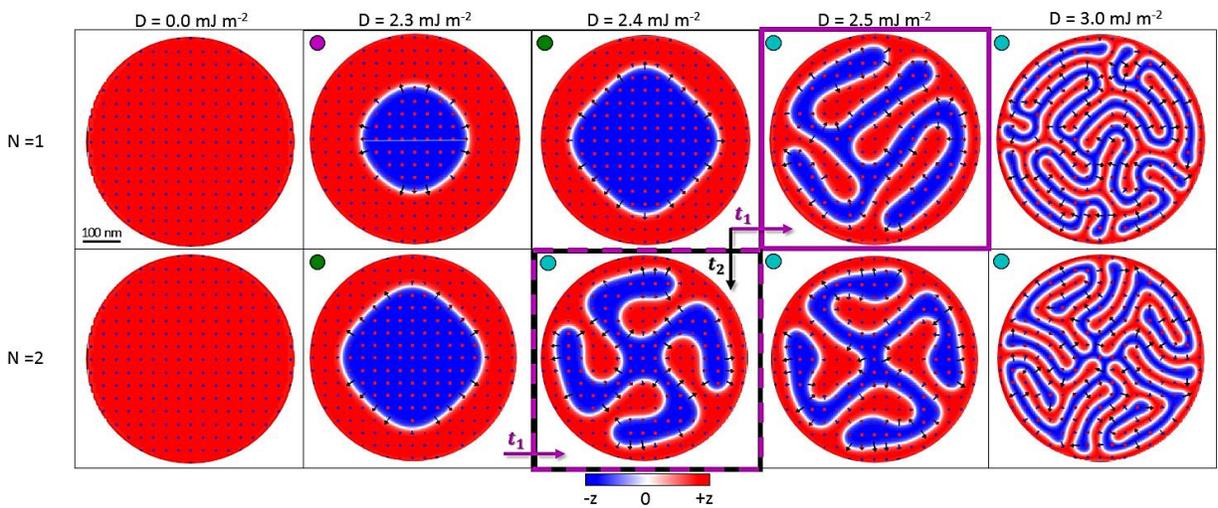

Figure 4 - Micromagnetic simulations for [ Pt \ Co \ Ir ] stacks of $N$ = 1 (top row) and $N$ = 2 (bottom row) in discs of $d_d$ = 500 nm for various DMI energies. For $D$ = 0.0 mJ m$^{-2}$ the initial skyrmion bubble collapsed onto itself, whilst when $D$ = 2.3 mJ m$^{-2}$ and $N$ =1 the bubble is stable in the disc. As can be seen at $D$ = 2.5 mJ m$^{-2}$ and $N$ = 1, a magnetic phase transition into a stable cycloidal state can occur by increasing the DMI energy, highlighted in magenta as a $t_1$ transition. This can also be accomplished by keeping DMI energy constant $D$ = 2.4 mJ m$^{-2}$ and increasing the dipolar energy by an additional layer repetition $N$ =2, highlighted in black as a $t_2$ transition. The colored circle marks in the upper left corner correspond to the skyrmion bubble (magenta), transitional state (green), and cycloidal state (light blue).

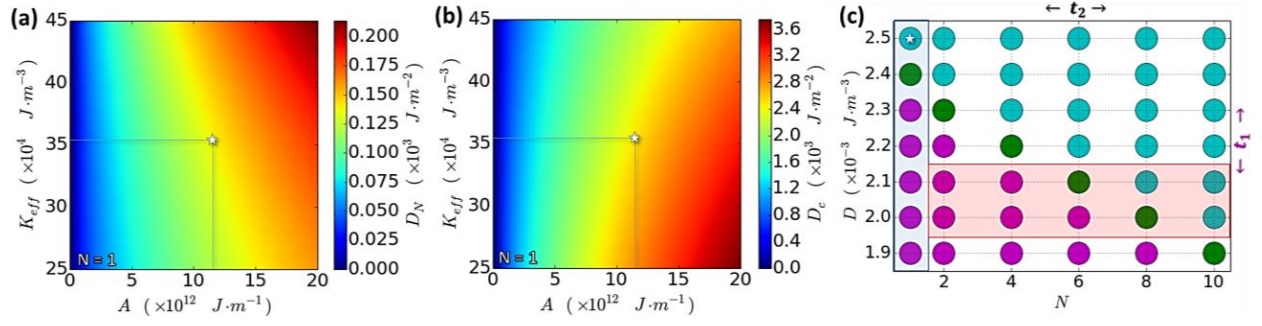

Figure 5 – DMI and layer repetitions. (a) The critical DMI energy needed to stabilize Néel walls ($D_N$) for $N = 1$ as a function of the $A$ and $K_{eff}$ for $t = 0.8$ nm and $M_s = 0.98$ MA m$^{-1}$. (b) The critical DMI energy needed to stabilize a cycloidal state ($D_c$) for $N = 1$ as a function of $A$ and $K_{eff}$ under similar constraints. (c) A simulated phase diagram as a function of [ Pt \ Co \ Ir ] layer repetitions $N$ and DMI energy $D$ for $d_d = 500$ nm discs. The colored markers correspond to the skyrmion bubbles (magenta), transitional (green), and cycloidal (light blue) states shown in Figure 4. For $N = 1$, the numeric approximation of $D_c = 2.5$ mJ m$^{-2}$ agrees with the analytical value (star marker). From experimental observations our system transitions into a cycloidal state for $N = 10$ which would suggest a minimum DMI energy of $D = 2.0$ mJ m$^{-2}$ outlined by the red box overlay.